\documentstyle[12pt]{article}

\newcommand{\be}{\begin{equation}}
\newcommand{\ee}{\end{equation}}
\newcommand{\bea}{\begin{eqnarray}}
\newcommand{\eea}{\end{eqnarray}}
\newcommand{\beaa}{\begin{eqnarray*}}
\newcommand{\eeaa}{\end{eqnarray*}}

\newcommand{\BB}{{{\rm I} \kern -2pt \rlap {\rm B} \kern +8pt}}

\advance\textheight by \topskip
\begin{document}

\baselineskip 18pt \parindent 12pt \parskip 10pt

\begin{titlepage}

\begin{center}
{\Large {\bf Conserved Quantities in Noncommutative Principal
Chiral Model with Wess-Zumino Term }}\\\vspace{1.5in} {\large
U. Saleem\footnote{%
usman\_physics@yahoo.com}, M. Hassan \footnote{%
mhassan@physics.pu.edu.pk} }\vspace{0.15in} and M. Siddiq\footnote{%
mohsin\_pu@yahoo.com}\footnote{%
On study leave from PRD (PINSTECH) Islamabad, Pakistan}

{\small{\it Department of Physics,\\ University of the Punjab,\\
Quaid-e-Azam Campus,\\Lahore-54590, Pakistan.}}
\end{center}

\vspace{1cm}
\begin{abstract}
We construct noncommutative extension of $U(N)$ principal chiral
model with Wess-Zumino term and obtain an infinite set of local
and non-local conserved quantities for the model using iterative
procedure of Brezin {\it et.al} \cite{BIZZ}. We also present the
equivalent description as Lax formalism of the model. We expand
the fields perturbatively and derive zeroth- and first-order
equations of motion, zero-curvature condition, iteration method,
Lax formalism, local and non-local conserved quantities.
\end{abstract}
\vspace{1cm} PACS: 11.10.Nx, 02.30.Ik
\end{titlepage}

During the last twenty years, the subject of classical and quantum
integrability of field theoretic models has attracted a great
interest \cite {Bab}-\cite{BIZZ}. The integrability of a field
theoretic model is guaranteed by the existence of an infinite
number of local and non-local conserved quantities
\cite{Pohl}-\cite{BIZZ}. These conserved quantities appear due to
the fact that the field equations of these models can be expressed
as zero-curvature condition for a given connection.

There is an increasing interest in noncommutative geometry
\cite{conne} due to its applications in particle physics, string
theory, differential geometry etc \cite{N1}-\cite{se}. For the
last few years, there has been an increasing interest in
noncommutative\footnote{%
The noncommutative spaces are defined by the noncommutativity of
the coordinates i.e.
\[
\lbrack x^{i},x^{j}]=\mbox{i}\theta ^{ij},
\]
where $\theta ^{ij}$ known as deformation parameters are real
constants.} extension of the field theories \cite{Fur}-\cite{moh}.
The noncommutative version of the integrable field theoretic
models is obtained by replacing ordinary product of fields in the
action by their $\star $-product \cite{moyal}%
\footnote{%
The product of two functions in noncommutative space is defined as
\[
\left( f\star g\right) (x)=f(x)g(x)+\frac{\mbox{i}\theta
^{ij}}{2}\partial _{i}f(x)\partial _{j}g(x)+\vartheta (\theta
^{2}),
\]
The following properties holds in noncommutative space
\begin{eqnarray*}
(f\star g)\star h &=&f\star (g\star h), \\
f\star I &=&f=I\star f.
\end{eqnarray*}
}. This has been shown that in general the noncommutativity of
time variables leads to non-unitarity and affects the causality of
the theory \cite{Gomis, se}. These noncommutative integrable
models reduce to ordinary models in the limit when the deformation
parameter $\theta$ vanishes. The noncommutative extension of
integrable models such as principal chiral model with and without
a Wess-Zumino term, Liouville equation, sine-Gordon equation,
Korteweg de Vries (KdV) equation, Boussinesq equation, Kadomtsev
Petviashvili (KP) equation, Sawada-Kotera equation, nonlinear
Schrodringer equation and Burgers equation have been studied by
different authors by exploiting different techniques
\cite{Fur}-\cite{Muller4}. The noncommutative sigma model
especially $U(N)$ principal chiral model has been investigated in
\cite{Marco1, Profumo} and some results regarding their
integrability have been obtained. The ultraviolet property of
Wess-Zumino-Witten (WZW) model has been studied in \cite{Fur}.

In this letter we present noncommutative extension of $U(N)$
principal chiral model with Wess-Zumino term (nc-PCWZM). We extend
iterative procedure of Brezin {\it et.al} \cite{BIZZ} for the
nc-PCWZM and derive a set of associated linear differential
equations. It has been shown that the set of linear differential
equations is equivalent to a noncommutative Lax formalism of the
model. We derive a series of local and non-local conserved
quantities. We expand the fields perturbatively and obtain zeroth-
and first-order equations of motion, zero-curvature condition, Lax
formalism, iterative procedure of Brezin {\it et.al}, local and
non-local conserved quantities.

The action for principal chiral model with Wess-Zumino term is
\cite{witten}
\[
\frac{\beta }{2}\int d^{2}x\mbox{Tr}(\partial _{+}g^{-1}\partial _{-}g)+%
\frac{\beta }{3}\kappa \int d^{2}x\mbox{Tr}(g^{-1}dg)^{3},
\]
with constraint \thinspace \thinspace \thinspace \thinspace
\thinspace \thinspace \thinspace \thinspace \thinspace \thinspace
\thinspace \thinspace \thinspace
\[
gg^{-1}=g^{-1}g=1.
\]
This is referred to as principal chiral Wess-Zumino model (PCWZM).
The constants $\beta $ and $\kappa $ are dimensionless and the
field $g(x^{\pm })$ is function of space-time coordinate $x^{\pm
}$. We define conserved currents of PCWZM associated with the
global symmetry of the model, as
\[
\bar{j}_{\pm }\equiv(1\pm \kappa )j_{\pm }=-(1\pm \kappa
)g^{-1}\partial _{\pm }g.
\]
The equation of motion and zero-curvature condition for the PCWZM
are
\begin{eqnarray*}
\partial _{-}\,\bar{j}_{+}+\partial _{+}\,\bar{j}_{-} &=&0, \\
\partial _{-}\bar{j}_{+}-\partial _{+}\bar{j}_{-}+[\bar{j}_{+},\bar{j}_{-}]
&=&0,
\end{eqnarray*}
and the above two equations give
\[
\partial _{-}\,\bar{j}_{+}=-\frac{1}{2}[\bar{j}%
_{+},\bar{j}_{-}]=-\partial _{+}\,\bar{j}_{-},
\]
where $[\bar{j}_{+},\bar{j}_{-}]=\bar{j}%
_{+}\bar{j}_{-}-\bar{j}_{-}\bar{%
j}_{+}$ is commutator. The action for the principal chiral model
with Wess-Zumino term in
noncommutative space is obtained by replacing ordinary product of the fields with $\star$%
-product i.e.
\[
\frac{\beta }{2}\int d^{2}x\mbox{Tr}(\partial _{+}g^{-1}\star
\partial _{-}g)+\frac{\beta }{3}\kappa \int
d^{2}x\mbox{Tr}(g^{-1}\star dg)_{\star }^{3}\,,
\]
with constraint \thinspace \thinspace \thinspace \thinspace
\thinspace \thinspace \thinspace \thinspace \thinspace \thinspace
\thinspace \thinspace
\[
\,g\star g^{-1}=g^{-1}\star g=1.
\]
This is referred to as noncommutative principal chiral Wess-Zumino
model (nc-PCWZM). The field $g(x^{\pm })$ is function of
space-time coordinate $x^{\pm }$ in noncommutative space. The
conserved currents for nc-PCWZM are
\[
\bar{j}_{\pm }^{\star }=(1\pm \kappa )j_{\pm }^{\star }=-(1\pm
\kappa )g^{-1}\star \partial _{\pm }g.
\]
The current conservation equation is
\begin{equation}
\partial _{-}\bar{j}_{+}^{\star }+\partial _{+}\bar{j}_{-}^{\star }\,=\,0,
\label{motion}
\end{equation}
and the zero-curvature condition is
\begin{equation}
\partial _{-}\bar{j}_{+}^{\star }-\partial _{+}\bar{j}_{-}^{\star }+[\bar{j}%
_{+}^{\star }\,,\,\bar{j}_{-}^{\star }]_{\star }=\,0,
\label{zero}
\end{equation}
where $[\bar{j}_{+}^{\star }\,,\,\bar{j}_{-}^{\star }]_{\star }=\bar{j}%
_{+}^{\star }\,\star \,\bar{j}_{-}^{\star }-\,\bar{j}_{-}^{\star }\star \bar{%
j}_{+}^{\star }$ is commutator in noncommutative space. The
noncommutative equations (\ref {motion}) and (\ref{zero}) can be
combined to give
\begin{equation}
\partial _{-}\bar{j}_{+}^{\star }=-\frac{1}{2}[\bar{j}_{+}^{\star }\,,\,\bar{%
j}_{-}^{\star }]_{\star }=-\partial _{+}\bar{j}_{-}^{\star }.
\label{zero1}
\end{equation}
It is straightforward to derive an infinite set of local conserved
quantities from equation (\ref{zero1}). We get a series of local
conserved quantities
\begin{equation}
\partial _{\mp }\mbox{Tr}\left( \bar{j}_{\pm }^{\star }\right)_{\star}
^{n}=0,
\label{localseries}
\end{equation}
where the numbers $n$ are precisely the exponents of the Lie
algebra of $U(N)$. These conserved quantities are local functions
of fields. The term local in this context is used for those
conserved currents which depend on fields and their derivatives
but not on their integrals. This should not be confused with the
intrinsic non-locality of a noncommutative field theory emerging
due to derivatives of fields to an infinite order, multiplied with
the deformation parameter $\theta$. This intrinsic non-locality
due to noncommutativity persists in all definitions of conserved
currents. In the ordinary commutative case, these conserved
currents can be used to construct quantities which are in
involution with each other and there spins are exponents of the
Lie algebra modulo the coxter number \cite{Hassan1}. It is
however, not straightforward to see in the noncommutative case,
whether the conserved quantities obtained from equation
(\ref{localseries}) are in involution or not. The involution of
the local conserved quantities can only be established, if we know
the Poisson bracket current algebra of nc-PCWZM. The canonical
formalism is quite involved and does not allow simply to write the
noncommutative Poisson current algebra of the model.

In nc-PCWZM, we also encounter with an infinite number of another
type of conserved currents which are non-local as they depend
non-locally on fields i.e. they depend on fields, their
derivatives and their integrals as well. In order to derive the
non-local conserved currents, we define covariant derivative in
noncommutative space-time acting on some scalar field $\chi$, such
that
\begin{equation} D_{\pm }\chi=(1\pm \kappa )(\partial _{\pm
}\chi -j_{\pm }^{\star }\star \chi )\qquad \qquad \Rightarrow
\qquad \qquad [D_{+}\,,\,D_{-}]_{\star }\,=\,0, \label{cov}
\end{equation}
where $[D_{+}\,,\,D_{-}]_{\star }=D_{+}\,\star
\,D_{-}-D_{-}\,\star \,D_{+}.\,$\thinspace Now we suppose that
there exist currents $\bar{j}_{\pm }^{\star (k)}$ for $k=1,2,\dots
,n$ which are conserved
\begin{equation}
\partial _{-}\bar{j}_{+}^{\star (k)}+\partial _{+}\bar{j}_{-}^{\star
(k)}\,=\,0,  \label{para}
\end{equation}
such that
\begin{equation}
\bar{j}_{\pm }^{\star (k)}=\pm \partial _{\pm }\chi%
^{(k)}.  \label{para1}
\end{equation}
Further we have define $(k+1)$th current as
\begin{eqnarray}
\bar{j}_{\pm }^{\star (k+1)} &=&D_{\pm }\chi^{(k)}  \label{deri} \\
&=&\,(1\pm \kappa )\partial _{\pm }\chi ^{(k)}-\bar{j}_{\pm
}^{\star }\star \chi ^{(k)}.  \nonumber
\end{eqnarray}
The current $\bar{j}_{\pm }^{\star (k+1)}$ is also conserved which
can be checked as follows
\begin{eqnarray*}
\partial _{-}\bar{j}_{+}^{\star (k+1)}+\partial _{+}\bar{j}_{-}^{\star
(k+1)} &=&(\partial _{-}\star D_{+}+\partial _{+}\star D_{-})\chi^{(k)} \\
&=&(D_{+}\star \partial _{-}+D_{-}\star \partial _{+})\chi^{(k)} \\
&=&-D_{+}\star \bar{j}_{- }^{\star (k)}+D_{-}\star \bar{j}_{+}^{\star (k)}\\
&=&-[D_{+}, D_{-}]_{\star }\chi^{(k-1)} \\
&=&0,
\end{eqnarray*}
where $D_{\pm }\star \partial _{\pm }=\partial _{\pm }\star D_{\pm }$. We set $%
\bar{j}_{\pm }^{\star (1)}=\bar{j}_{\pm }^{\star }$ ,
$\bar{j}_{\pm }^{\star (0)}=0$ and $\chi^{(0)}=1$. Note that the
conservation of $k$th current implies the conservation of
$(k+1)$th current and as a result an infinite number of conserved
currents are obtained through induction.

The non-local conserved quantities in noncommutative space are now
defined as
\begin{equation}
\bar{Q}^{\star (k)}\,=\,\int_{-\infty }^{\infty
}dy\,\bar{j}_{0}^{\star (k)}(t,y). \label{charge}
\end{equation}
From equations (\ref{para1}) and (\ref{deri}), we get
\begin{eqnarray*}
\,\bar{j}_{0}^{\star (k+1)} &\equiv&\frac{1}{2}\left( \bar{j}_{+}^{\star (k+1)}+%
\bar{j}_{-}^{\star (k+1)}\right) =\partial _{0}\chi^{(k)}+\kappa
\partial _{1}\chi^{(k)}-\,\bar{j}_{0}^{\star }\star \chi^{(k)},
\\
\bar{j}_{1}^{\star (k+1)} & \equiv &\frac{1}{2}\left( \bar{j}_{+}^{\star (k+1)}-\bar{j}%
_{-}^{\star (k+1)}\right)=\partial _{1}\chi^{(k)}+\kappa \partial _{0}%
\chi^{(k)}-\,\bar{j}_{1}^{\star }\star \chi^{(k)},
\end{eqnarray*}
where
\[
\chi^{(k)}(t,y)=\,\int_{-\infty }^{y}dz\,\bar{j}_{0}^{\star
(k)}(t,z).
\]
The second current is a non-local current, it depends non-locally
on fields,
\begin{equation}
\bar{j}_{0}^{\star (2)}(t,y)=\bar{j}_{1}^{\star }(t,y)+\kappa \,\bar{j}%
_{0}^{\star }(t,y)-\bar{j}_{0}^{\star }(t,y)\star \int_{-\infty }^{y}dz\,\bar{j}%
_{0}^{\star }(t,z).  \label{sec}
\end{equation}
The first two conserved charges are
\begin{eqnarray}
\bar{Q}^{\star (1)}\, &=&\,\int dy\,\bar{j}_{0}^{\star }(t,y),  \nonumber \\
\bar{Q}^{\star (2)} &=&\,\int dy\left( \bar{j}_{1}^{\star }(t,y)+\kappa \bar{j}%
_{0}^{\star }(t,y)-\bar{j}_{0}^{\star }(t,y)\star \int_{-\infty }^{y}dz\,\bar{j}%
_{0}^{\star }(t,z)\right) .  \label{charges}
\end{eqnarray}
These conserved quantities reduce to the conserved quantities
obtained in Ref.\cite{Abb} in the limit when the deformation
parameter approaches to zero. In the limit when $\kappa
\rightarrow 0$ the conserved quantities (\ref{charges}) reduce to
the conserved quantities obtained for noncommutative principal
chiral model (nc-PCM) in Ref.\cite{Marco1}.

We now relate the iterative procedure outlined above to the
noncommutative Lax formalism of nc-PCWZM. From equations
(\ref{para1}) and (\ref{deri}), we have
\[\partial_{\pm}\chi^{(k)} =\pm D_{\pm}\chi^{(k-1)}.
\]
Multiplying by $\lambda^k$ and summing from $k=1$ to $k=\infty$,
we write
\[\sum_{k=1}^{\infty}(\lambda^k\partial_{\pm}\chi^{(k)}
)=\pm \sum_{k=1}^{\infty}(\lambda^kD_{\pm}\chi^{(k-1)}).
\]As $\chi^{(0)}=1$, the summation on right hand side can be extended
to $k=0$. The associated linear system for the nc-PCWZM can be
written as
\begin{eqnarray}
\partial_{\pm}u(t,x;\lambda )&=&\bar{A}_{\pm}^{\star(\lambda)}\star u(t,x;\lambda ),\label{linear1}
\end{eqnarray}
where the noncommutative fields $\bar{A}_{\pm}^{\star(\lambda)}$
are
\[\bar{A}_{+}^{\star(\lambda)}=\frac{-\lambda}{1-\lambda(1+\kappa)}
\bar{j}_{+}^{\star},\qquad
\bar{A}_{-}^{\star}=\frac{\lambda}{1+\lambda(1-\kappa)}\bar{j}_{-}^{\star},
\]
and $u(t,x;\lambda )$ is expanded as
\[
u(t,x;\lambda )=\sum_{k=0}^{\infty}\lambda^k\chi^{(k)}.
\]The compatibility condition of the linear system for the nc-PCWZM
is the $\star$-zero-curvature condition
\begin{eqnarray}
\left[ \partial _{+}-\bar{A}_{+}^{\star (\lambda
)},\partial_{-}-\bar{A}_{-}^{\star (\lambda )}\right] _{\star
}\equiv\partial_{-}\bar{A}_{+}^{\star(\lambda)}-\partial_{+}
\bar{A}_{-}^{\star(\lambda)}+[\bar{A}_{+}^{\star(\lambda)},
\bar{A}_{-}^{\star(\lambda)}]_{\star}=0.\label{zeronc}
\end{eqnarray} Now we define Lax operators for the nc-PCWZM,
\begin{eqnarray}
\bar{L}_{\pm}^{\star(\lambda)}=\partial_{\pm}-\bar{A}_{\pm}^{\star(\lambda)},\label{Lax1}
\end{eqnarray}
obeying the following equations
\[
\partial_{\mp}\bar{L}_{\pm}^{\star(\lambda)}=[\bar{A}_{\mp}^{\star(\lambda)},\bar{L}_{\pm}^{\star(\lambda)}]_{\star}
.\] The associated linear system (\ref{linear1}) can be
re-expressed in terms of space-time coordinates as
\begin{equation}
\partial _{0}u(t,x;\lambda)=\bar{A}_{0}^{\star (\lambda )}\star u(t,x;\lambda
),\qquad \partial _{1}u(t,x;\lambda )=\bar{A}_{1}^{\star (\lambda
)}\star u(t,x;\lambda ),\label{dual1}
\end{equation}
with the noncommutative fields $\bar{A}_{0}^{\star (\lambda )}$
and $\bar{A}_{1}^{\star (\lambda )}$ given by
\[
\bar{A}_{0}^{\star (\lambda )}=\frac{1}{2}\left(
\bar{A}_{+}^{\star (\lambda )}+\bar{A}_{-}^{\star(\lambda
)}\right),\qquad \bar{A}_{1}^{\star (\lambda )}=\frac{1}{2}\left(
\bar{A}_{+}^{\star(\lambda ) }-\bar{A}_{-}^{\star(\lambda
)}\right).
\]
The associated Lax operator $\bar{L}_{0}^{\star (\lambda
)}=\partial _{0}-\bar{A}_{0}^{\star (\lambda )}$, obeys the
following Lax equation
\[
\partial _{0}\bar{L}_{1}^{\star (\lambda )}=[\bar{A}_{0}^{\star (\lambda )},\bar{L}_{1}^{\star
(\lambda )}]_{\star }.
\]
This gives the time involution of the Lax operator related to the
isospectral problem of the model. We note that the Lax formalism
can be generalized to its noncommutative version without any
constraints.

Let $\hat{g}$ be a solution of equation of motion, therefore, the
associated linear system for the fields $\hat{g}$ can be written
as
\[
\partial _{\pm }\hat{u}(t,x;\lambda )=\hat{\bar{A}}_{\pm }^{\star (\lambda
)}\star\hat{u}(t,x;\lambda ),
\]
where the noncommutative fields $\hat{\bar{A}}_{\pm }^{\star
(\lambda )}$ are given by
\[\hat{\bar{A}}_{+}^{\star(\lambda)}=\frac{-\lambda}{1-\lambda(1+\kappa)}\hat{\bar{j}}_{+}^{\star},\qquad
\hat{\bar{A}}_{-}^{\star}=\frac{\lambda}{1+\lambda(1-\kappa)}\hat{\bar{j}}_{-}^{\star}
.\] The compatibility condition of the above linear system is
\begin{equation}
\left[ \partial _{+}-\hat{\bar{A}}_{+}^{\star (\lambda )},\partial _{-}-\hat{\bar{A}}%
_{-}^{\star (\lambda )}\right] _{\star }\equiv \partial _{-}\hat{\bar{A}}%
_{+}^{\star (\lambda )}-\partial _{+}\hat{\bar{A}}_{-}^{\star
(\lambda )}+\left[\hat{\bar{A}}_{+}^{\star (\lambda
)},\hat{\bar{A}}_{-}^{\star (\lambda )}\right] _{\star }=0.
\label{dar2}
\end{equation}
The Darboux transformation relates the matrix function  $\hat{u}%
(t,x;\lambda )$ and  $u(t,x;\lambda )$ by
\begin{equation}
\hat{u}=u\star v,  \label{dar1}
\end{equation}
and one can easily get
\begin{equation}
\partial _{\pm }v=\hat{\bar{A}}_{\pm }^{\star(\lambda )}\star v-v\star \bar{A}_{\pm }^{\star(\lambda )}.  \label{dar}
\end{equation}
The equation (\ref {dar}) is also known as B\"{a}cklund
transformation.

An infinite number of non-local conserved quantities can also be
generated from the Lax formalism of nc-PCWZM. We assume the
spatial
boundary conditions such that the currents $\bar{j}^{\star}_{\pm}$ vanish as $%
x\rightarrow \pm \infty $. The equation (\ref{dual1}) implies that $%
u(t,\infty ;\lambda )$ are time independent. The residual freedom
in the solution for $u(t,\infty ;\lambda )$ allows us to fix
$u(t,\infty ;\lambda )=1$, the unit matrix and we are then left
with time-independent matrix valued function
\begin{equation}
\bar{Q}^{\star}(\lambda )=u(t,\infty ;\lambda ).  \label{ch}
\end{equation}
Expanding $\bar{Q}^{\star}(\lambda )$ as power series in $\lambda
$, gives an infinite number of non-local conserved quantities
\begin{equation}
\bar{Q}^{\star}(\lambda )=\sum_{k=o}^{\infty }\lambda ^{k}\bar{Q}^{\star (k)}\qquad ,\qquad \qquad {%
\frac{d}{dt}}\bar{Q}^{\star (k)}=0.  \label{expn1}
\end{equation}
For the explicit expressions of the non-local conserved
quantities, we write (\ref {dual1}) as
\begin{equation}
u(t,x;{\lambda })=1-\frac{1}{2}\int_{-\infty }^{x}dy\,\left(
\bar{A}_{+}^{\star (\lambda )}(t,y)- \bar{A}_{-}^{\star(\lambda )
}(t,y)\right) \star u(t,y;{\lambda }).  \label{int}
\end{equation}
When we expand the field $u(t,x;{\lambda })$ as power series in
$\lambda $ as
\begin{equation}
u(t,x;{\lambda })=\sum_{k=o}^{\infty }\lambda ^{k}u_{k}(t,x),
\label{expn}
\end{equation}
and compare the coefficients of powers of $\lambda $, we get a
series of conserved non-local currents, which upon integration
give non-local conserved quantities (\ref{charges}). In the
commutative case, the local conserved quantities based on the
invariant tensors, all Poisson commute with each other and with
the non-local conserved quantities. The classical Poisson brackets
of non-local conserved quantities constitute classical Yangian
symmetry \cite{Hassan1}. In the noncommutative case we expect
similar results but once we know the Poisson bracket algebra of
the currents $\bar{j}_{\pm}^{\star}$, we shall be able to address
the Poisson bracket algebra of these conserved quantities.

We now study the perturbative expansion of the field $g(x^{\pm})$
and compute the equation of motion and zero-curvature condition,
Lax equation and the conserved quantities upto first order in the
deformation parameter $\theta $. We define group-valued field
$g(x^{\pm})$ by \cite{moh}
\[
g=\exp _{\star }\left( \frac{\mbox{i}\varphi }{2}\right) ,\qquad \mbox{and}%
\qquad g^{-1}=\exp _{\star }\left( -\frac{\mbox{i}\varphi
}{2}\right) .
\]
with
\[
g=\exp _{\star }\left( \frac{\mbox{i}\varphi }{2}\right) =1+\frac{\mbox{i}}{%
1!}\left( \frac{\varphi }{2}\right) +\frac{1}{2!}\left( \frac{%
\mbox{i}}{2}\right)^{2}\varphi \star\varphi + \dots ,
\]
where $\varphi$ is in the Lie algebra of $U(N).$ The components of
Noether's currents are
\[
\bar{j}_{\pm }^{\star }=-(1\pm \kappa )g^{-1}\star \partial _{\pm
}g,
\]
or
\begin{equation}
\bar{j}_{\pm }^{\star }=-(1\pm \kappa )\frac{\mbox{i}}{2}\partial
_{\pm
}\varphi -(1\pm \kappa )^{2}\frac{\theta }{2!}\left( \frac{\mbox{i}}{2}%
\right) ^{2}\left( \partial _{\pm }^{2}\varphi \partial _{\mp
}\varphi -\partial _{\mp }\partial _{\pm }\varphi \partial _{\pm
}\varphi \right) +\vartheta (\theta ^{2}).  \label{current}
\end{equation}
We expand $\varphi $ as a power series in $\theta $
\begin{equation}
\varphi =\varphi ^{[0]}+\theta \varphi ^{[1]},\qquad \bar{j}_{\pm }^{\star }=%
\bar{j}_{\pm }^{[0]}+\theta \,\,\tilde{\bar{j}}_{\pm }^{[1]}
\label{current1}
\end{equation}
where
\begin{eqnarray*}
\bar{j}_{\pm }^{[0]} &=&-\frac{\mbox{i}}{2}(1\pm \kappa )\partial
_{\pm }\varphi ^{[0]},\qquad \bar{j}_{\pm
}^{[1]}=-\frac{\mbox{i}}{2}(1\pm \kappa )\partial _{\pm }\varphi
^{[1]}, \\
\tilde{\bar{j}}_{\pm }^{[1]} &=&\bar{j}_{\pm }^{[1]}-\frac{1}{2!}\left( \bar{j}%
_{\pm \pm }^{[0]}\bar{j}_{\mp }^{[0]}-\frac{1}{2}\bar{j}_{\mp
}^{[0]}\bar{j}_{\pm }^{[0]^{2}}+\frac{1}{2}\bar{j}_{\pm
}^{[0]}\bar{j}_{\mp }^{[0]}\bar{j}_{\pm }^{[0]}\right) .
\end{eqnarray*}
By substituting the value of $\bar{j}_{\pm }^{\star }$ from
equation (\ref {current1}) in equations of motion (\ref{motion})
and (\ref{zero}), we get
\begin{eqnarray*}
\partial _{-}\bar{j}_{+}^{[0]}+\partial _{+}\bar{j}_{-}^{[0]} &=&0, \\
\partial _{-}\tilde{\bar{j}}_{+}^{[1]}+\partial _{+}\tilde{\bar{j}}_{-}^{[1]} &=&0, \\
\partial _{-}\bar{j}_{+}^{[0]}-\partial _{+}\bar{j}_{-}^{[0]}+\left[ \bar{j}%
_{+}^{[0]},\bar{j}_{-}^{[0]}\right]  &=&0, \\
\partial _{-}\,\tilde{\bar{j}}_{+}^{[1]}-\partial _{+}\,\tilde{\bar{j}}_{-}^{[1]}+\left[ \,%
\tilde{\bar{j}}_{+}^{[1]},\bar{j}_{-}^{[0]}\right] +\left[
\bar{j}_{+}^{[0]},\,\tilde{\bar{j}}%
_{-}^{[1]}\right]  &=&-\frac{\mbox{i}}{2}\left( \bar{j}_{++}^{[0]}\bar{j}%
_{--}^{[0]}+\bar{j}_{--}^{[0]}\bar{j}_{++}^{[0]}\right) -
\frac{\mbox{i}}{8}\left[ \bar{j}%
_{+}^{[0]}, \bar{j}_{-}^{[0]}\right]^2 .
\end{eqnarray*}
It is clear from above equations that currents are conserved upto
first order in $\theta $ but zero-curvature condition is spoiled.
The perturbative expansion of iterative construction is
\begin{eqnarray*} \bar{j}_{\pm}^{[0](k+1)}&=&D_{\pm
}^{[0]}\chi^{[0](k)},\qquad\qquad\qquad\,\,\,\,\,\,\Rightarrow
\qquad
\partial _{-}\bar{j}_{+}^{[0](k+1)}+\partial _{+}\bar{j}_{-}^{[0](k+1)}=0, \\\tilde{\bar{j}}_{\pm }^{[1](k+1)}&=&D_{\pm
}^{[0]}\chi^{[1](k)}-\tilde{D}_{\pm }^{[1]}\chi^{[0](k)},\qquad
\Rightarrow \qquad
\partial _{-}\tilde{\bar{j}}_{+}^{[1](k+1)}+\partial
_{+}\tilde{\bar{j}}_{-}^{[1](k+1)}=0,
\end{eqnarray*}
where
\begin{eqnarray*}
D_{\pm}^{[0]}\chi^{[1](k)}=\partial_{\pm}\chi^{[1](k)}-\bar{j}_{\pm}^{[0]} \chi^{[1](k)},\\
\tilde{D}_{\pm}^{[1]}\chi^{[0](k)}=\tilde{\bar{j}}_{\pm}^{[1]}\chi^{[0](k)}+\frac{\mbox{i}}{2}%
(\partial_{\pm}\bar{j}_{\pm}^{[0]}\partial_{\mp}-\partial_{\mp}\bar{j}_{\pm}^{[0]}\partial_{\pm})\chi^{[0](k)}.
\end{eqnarray*}
From this analysis, it is obvious that the conservation of $k$th
current implies the conservation of $(k+1)$th current upto first
order in $\theta$.

The zero-curvature condition (\ref{zeronc}) for the fields
$\bar{A}_{\pm}^{\star(\lambda)}$ in perturbative expansion becomes
\begin{eqnarray}
&&\left. \partial _{-}\bar{A}_{+}^{[0](\lambda )}-\partial
_{+}\bar{A}_{-}^{[0](\lambda )}+\left[ \bar{A}_{+}^{[0](\lambda
)},\bar{A}_{-}^{[0](\lambda )}\right] =0,\right.
\label{0Lax1} \\
&&\left. \partial _{-}\,\tilde{\bar{A}}_{+}^{[1](\lambda
)}-\partial _{+}\,\tilde{\bar{A}}_{-}^{[1](\lambda )}+\left[
\tilde{\bar{A}}_{+}^{[1](\lambda
)},\bar{A}_{-}^{[0](\lambda )}\right] +\left[ \bar{A}_{+}^{[0]},\,\tilde{\bar{A}}%
_{-}^{[1](\lambda )}\right] =-\frac{\mbox{i}}{2}\left(
\bar{A}_{++}^{[0](\lambda )}\bar{A}_{--}^{[0](\lambda
)}+\bar{A}_{--}^{[0](\lambda )}\bar{A}_{++}^{[0](\lambda )}\right)\right.+   \nonumber \\
&&\left. \qquad\qquad\qquad\qquad \qquad\qquad\qquad
+\frac{\mbox{i}}{2}\left(
\partial _{-}\bar{A}_{+}^{[0](\lambda )}\partial _{+}\bar{A}_{-}^{[0](\lambda
)}+\partial _{+}\bar{A}_{-}^{[0](\lambda )}\partial
_{-}\bar{A}_{+}^{[0](\lambda )}\right) ,\right.   \label{1Lax2}
\end{eqnarray}
where the fields $\bar{A}_{\pm }^{[0](\lambda )}$ and
$\tilde{\bar{A}}_{\pm }^{[1](\lambda )}$ are given by
\begin{eqnarray*}
\bar{A}_{+}^{[0](\lambda)}=\frac{-\lambda}{1-\lambda(1+\kappa)}\bar{j}_{+}^{[0]},\qquad
\bar{A}_{-}^{[0]}=\frac{\lambda}{1+\lambda(1-\kappa)}\bar{j}_{-}^{[0]},\\
\tilde{\bar{A}}_{+}^{[1](\lambda)}=\frac{-\lambda}{1-\lambda(1+\kappa)}\tilde{\bar{j}}_{+}^{[1]},\qquad
\tilde{\bar{A}}_{-}^{[1]}=\frac{\lambda}{1+\lambda(1-\kappa)}\tilde{\bar{j}}_{-}^{[1]}.
\end{eqnarray*}
The perturbative expansion of the linear system (\ref{linear1}) is
\begin{eqnarray*}
\partial_{\pm}u^{[0]}(x,t;\lambda)&=&\bar{A}_{\pm }^{[0](\lambda)}u^{[0]}(x,t;\lambda), \\
\partial_{\pm}u^{[1]}(x,t;\lambda)&=&\bar{A}_{\pm }^{[0](\lambda
)}u^{[1]}(x,t;\lambda)+(\tilde{\bar{A}}_{\pm
}^{[1](\lambda)}+\frac{\mbox{i}}{2}(\bar{A}_{\pm\pm }^{[0](\lambda
)}\bar{A}_{\mp }^{[0](\lambda)}-\bar{A}_{\pm\mp}^{[0](\lambda
)}\bar{A}_{\pm }^{[0](\lambda )}))u^{[0]}(x,t;\lambda).
\end{eqnarray*}
The compatibility conditions of first and second equations in the
above set of equations are (\ref{0Lax1}) and (\ref{1Lax2}),
respectively. The perturbative expansion of the Lax operators
(\ref{Lax1}) is expressed as
\[
\bar{L}_{\pm }^{(\lambda )}=\bar{L}_{\pm }^{[0](\lambda
)}+\theta\tilde{\bar{L}}_{\pm }^{[1](\lambda )},
\]
where $\bar{L}_{\pm }^{[0](\lambda )}$ and $\tilde{\bar{L}}_{\pm
}^{[1](\lambda )}$ are given by
\begin{eqnarray*}
\bar{L}_{\pm }^{[0](\lambda )} &=&\partial _{\pm }-\bar{A}_{\pm }^{[0](\lambda )}, \\
\tilde{\bar{L}}_{\pm }^{[1](\lambda )} &=&-\tilde{\bar{A}}_{\pm
}^{[1](\lambda )}.
\end{eqnarray*}
obeying the following equations
\begin{eqnarray}
\partial _{\mp }\bar{L}_{\pm }^{[0](\lambda )} &=&[\bar{A}_{\mp }^{[0](\lambda
)}\,,\bar{L}_{\pm }^{[0](\lambda )}],  \label{PLAX1} \\
\partial _{\mp }\tilde{\bar{L}}_{\pm }^{[1](\lambda )} &=&[\bar{A}_{\mp }^{[0](\lambda
)}\,,\tilde{\bar{L}}_{\pm }^{[1](\lambda )}]+[\tilde{\bar{A}}_{\mp
}^{[1](\lambda )}\,,\bar{L}_{\pm }^{[0](\lambda
)}]-\frac{\mbox{i}}{2}\left( \bar{L}_{\pm\pm}^{[0](\lambda
)}\bar{A}_{\mp\mp}^{[0](\lambda )}+\bar{A}_{\mp\mp}^{[0](\lambda
)}\bar{L}_{\pm\pm}^{[0](\lambda )}\right)
\nonumber \\
&&+\frac{\mbox{i}}{2}\left( \partial
_{\mp}\bar{L}_{\pm}^{[0](\lambda )}\partial
_{\pm}\bar{A}_{\mp}^{[0](\lambda )}+\partial
_{\pm}\bar{A}_{\mp}^{[0](\lambda )}\partial
_{\mp}\bar{L}_{\pm}^{[0](\lambda )}\right) .  \label{PLAX2}
\end{eqnarray}
The Lax equations (\ref{PLAX1}) and (\ref{PLAX2}) are equivalent to (\ref{0Lax1}) and (%
\ref{1Lax2}) respectively.

By substituting the value of $\bar{j}_{\pm }^{\star }$ from
equation (\ref {current1}) in equation (\ref{localseries}), we
find the following zeroth order and first order local conservation
laws,
\begin{eqnarray*}
&&\left. \partial _{\mp }{\rm Tr}\left( \bar{j}_{\pm
}^{[0]2}\right)
=0,\right.  \\
&&\left. \partial _{\mp }{\rm Tr}\left( \bar{j}_{\pm
}^{[0]}\tilde{\bar{j}}_{\pm}^{[1]}\right)
=0,\right.  \\
&&\left. \partial _{\mp }{\rm Tr}\left( \bar{j}_{\pm
}^{[0]3}\right)
=0,\right.  \\
&&\left. \partial _{\mp }{\rm Tr}\left( \bar{j}_{\pm
}^{[0]2}\tilde{\bar{j}}_{\pm}^{[1]}-\frac{1}{4}\bar{j}_{\pm }^{[0]}\bar{j}_{\pm \pm }^{[0]}\left[ \bar{j}%
_{\pm }^{[0]},\bar{j}_{\mp
}^{[0]}\right]+\frac{1}{4}\bar{j}_{\pm\pm
}^{[0]}\bar{j}_{\pm}^{[0]}[%
\bar{j}_{\pm }^{[0]},\bar{j}_{\mp}^{[0]}]\right) =0,\right.  \\
&&\left. \partial _{\mp }{\rm Tr}\left( \bar{j}_{\pm
}^{[0]4}\right)
=0,\right.  \\
&&\left. \partial _{\mp }{\rm Tr}\left( \bar{j}_{\pm
}^{[0]3}\tilde{\bar{j}}_{\pm}^{[1]}-\frac{1}{8}\bar{j}_{\pm }^{[0]2}\bar{j}_{\pm \pm }^{[0]}\left[ \bar{j%
}_{\pm }^{[0]},\bar{j}_{\mp
}^{[0]}\right]+\frac{1}{8}\bar{j}_{\pm\pm
}^{[0]}\bar{j}_{\pm }^{[0]2}\left[ \bar{j%
}_{\pm }^{[0]},\bar{j}_{\mp }^{[0]}\right] \right) =0.\right.
\end{eqnarray*}
Similarly we can expand iterative procedure perturbatively upto
first order in the deformation parameter $\theta $. The first two
non-local conserved currents are
\begin{eqnarray*}
\bar{j}_{0}^{(1)[0]}(t,y) &=&\bar{j}_{0}^{[0]}(t,y),\\
 \tilde{\bar{j}}_{0}^{(1)[1]}(t,y)&=&\tilde{\bar{j}}_{0}^{[1]}(t,y),\\
\bar{j}_{0}^{(2)[0]}(t,y)&=&\bar{j}_{1}^{[0]}(t,y)+\kappa\bar{j}_{0}^{[0]}(t,y)-\bar{j} _{0}^{[0]}(t,y)\int_{-\infty }^{y}\bar{j}_{0}^{[0]}(t,z)dz, \\
\tilde{ \bar{j}}_{0}^{(2)[1]}(t,y)&=&\tilde{\bar{j}}_{1}^{[1]}(t,y)+\kappa \tilde{\bar{j}}%
_{0}^{[1]}(t,y)-\tilde{\bar{j}}_{1}^{[1]}(t,y)\int_{-\infty
}^{y}\bar{j}_{0}^{[0]}\,(t,z)dz-\nonumber\\\qquad\qquad \qquad&&-
\bar{j}_{1}^{[0]}(t,y)\int_{-\infty
}^{y}\tilde{\bar{j}}_{0}^{[1]}\,(t,z)dz+ \mbox{Total Derivative},
\end{eqnarray*}
and corresponding conserved quantities are
\begin{eqnarray*}
\bar{Q}^{(1)[0]} &=&\int_{-\infty }^{\infty }\bar{j}_{0}^{[0]}(t,y)dy, \\
\tilde{\bar{Q}}^{(1)[1]} &=&\int_{-\infty }^{\infty }\tilde{\bar{j}}_{0}^{[1]}(t,y)dy, \\
\bar{Q}^{(2)[0]} &=&\int_{-\infty }^{\infty }\left( \bar{j}%
_{1}^{[0]}(t,y)+\kappa
\bar{j}_{0}^{[0]}(t,y)-\bar{j}_{0}^{[0]}(t,y)\int_{-\infty
}^{y}\bar{j}_{0}^{[0]}(t,z)dz\right) dy, \\
\tilde{\bar{Q}}^{(2)[1]} &=&\int_{-\infty }^{\infty }\left( \tilde{\bar{j}}%
_{1}^{[1]}(t,y)+\kappa \tilde{\bar{j}}_{0}^{[1]}(t,y)-\tilde{\bar{j}}_{1}^{[1]}(t,y)\int_{-\infty }^{y}%
\bar{j}_{0}^{[0]}\,(t,z)dz-\bar{j}_{1}^{[0]}(t,y)\int_{-\infty }^{y}\tilde{\bar{j}}%
_{0}^{[1]}\,(t,z)dz\right) dy.
\end{eqnarray*}
Note that the first order correction to the first non-local
conserved quantity is an integral of non-local function of the
fields.

In summary, we have extended $U(N)$ principal chiral model with
Wess-Zumino term in noncommutative space and have shown nc-PCWZM
preserves its integrability without any extra constraint. A set of
nontrivial local and non-local conserved quantities are calculated
for the nc-PCWZM. We check the integrability of the model at a
perturbative level and write the zeroth- and first-order equations
of motion, zero-curvature condition, local and non-local conserved
quantities. Similar noncommutative extension can also be
investigated for the supersymmetric principal chiral model and WZW
model. Another interesting direction to pursue is to investigate
$\star$-Poisson bracket algebra of the currents of nc-PCWZM and to
check the involution of local conserved quantities.

{\Large Acknowledgements}

The authors acknowledge the enabling role of the Higher Education
Commission Islamabad, Pakistan and appreciate its financial
support through ``Merit Scholarship Scheme for PhD studies in
Science \& Technology (200 Scholarships)''.

\end{document}